\documentclass[preprint2]{aastex}
\usepackage{psfig}
\include{epsf}
\newcommand{\be}{\begin{equation}}
\newcommand{\ee}{\end{equation}}
\newcommand{\ba}{\begin{eqnarray}}
\newcommand{\ea}{\end{eqnarray}}
\newcommand{\bsf}[1]{\mbox{\begin{bfseries}\textsf{{#1}}\end{bfseries}}}

\def\simless{\mathbin{\lower 3pt\hbox
   {$\rlap{\raise 5pt\hbox{$\char'074$}}\mathchar"7218$}}}
\def\simgreat{\mathbin{\lower 3pt\hbox
   {$\rlap{\raise 5pt\hbox{$\char'076$}}\mathchar"7218$}}}   

\shorttitle{Visible and dark matter in ER0047$-$2808}
\shortauthors{Dye \& Warren}

\begin{document}


\title{Decomposition of the visible and dark matter in the
Einstein ring 0047$-$2808, by semi--linear inversion }


\author{S. Dye and S.J. Warren}
\affil{Astrophysics Group, Blackett Lab., Imperial College, 
Prince Consort Road, London, SW7 2BW, U.K.}

\begin{abstract}

We measure the mass density profile of the lens galaxy in the Einstein
ring system 0047$-$2808 using our semi--linear inversion method
developed in an earlier paper. By introducing an adaptively gridded
source plane, we are able to eliminate the need for regularisation of
the inversion. This removes the problem of a poorly defined number of
degrees of freedom, encountered by inversion methods that employ
regularisation, and so allows a proper statistical comparison between
models. We confirm the results of \citet{wayth04}, that the source is
double, and that a power--law model gives a significantly better fit
that the singular isothermal ellipsoid model. We measure a slope
$\alpha=2.11\pm0.04$. We find, further, that a dual--component
constant $M/L$ baryonic + dark halo model gives a significantly better
fit than the power--law model, at the $99.7\%$ confidence level. The
inner logarithmic slope of the dark halo profile is found to be
$0.87^{+0.69}_{-0.61}$ (95\% CL), consistent with the predictions of
CDM simulations of structure formation. We determine an unevolved
B--band mass to light ratio for the baryons (only) of
$3.05^{+0.53}_{-0.90} \,h_{65}\,M_{\odot}/L_{B\odot}$ (95\% CL). This
is the first measurement of the baryonic $M/L$ of a single galaxy by
purely gravitational lens methods. The baryons account for
$65^{+10}_{-18}$\% (95\% CL) of the total projected mass, or, assuming
spherical symmetry, $84^{+12}_{-24}$\% (95\% CL) of the total
three--dimensional mass within the mean radius of 1.16''
($7.5h_{65}^{-1}$kpc) traced by the ring. Finally, at the level of
$>3\sigma$, we find that the halo mass is rounder than the baryonic
distribution and that the two components are offset in orientation
from one another.

\end{abstract}


\keywords{astrophysics; gravitational lensing; dark matter}

\section{Introduction}
\label{sec_intro}

The $\Lambda-$CDM model has been outstandingly successful in
explaining the growth of structure in the Universe, to the extent that
it has been argued that we should now treat the theory as established
\citep{binney04}. Attempts to falsify the theory have focused mainly
on the predicted mass profiles in the centres of galaxies. N--body
simulations have established that a simple formulation, 
$\rho(r) \propto (r/r_s)^{-\gamma}(1+r/r_s)^{\gamma-3}$, 
accurately describes the density profiles of dark--matter halos across
a wide range of length scales. At radii much smaller than the
characteristic scale $r_s$, the density profile is cuspy, of power law
form, $\rho(r)\propto r^{-\gamma}$, with values of $\gamma$ in the
range 1 (\citet{nfw96}, the `NFW profile') to 1.5
\citep{moore98,moore99} indicated. However, significantly shallower
slopes than predicted have been inferred from observations of the
rotation curves of low surface brightness galaxies (LSBs)
\citep{deblok97,deblok01}. Because of this \citet{spergel00} argued
that the collisionless CDM picture requires modification, and that the
particles are self--interacting with a large scattering cross section.

More recent work indicates that these conclusions may be premature,
and at present the situation is unclear. Even with data of improved
spatial resolution, \citet{swaters03} emphasise sources of systematic
error in the measurement of $\gamma$ from galaxy rotation curves. They
find that their sample of 15 dwarf and LSB galaxy rotation curves does
not preclude a slope $\gamma=1$. At the same time, the most recent
simulations show that the central mass profiles do not reach an
asymptotic value of the slope, but that the slope continues to flatten
toward the centre, down to the resolution limit of the simulations
\citep{power03,navarro04}. This means that at the small radii of the
observations, $\sim1\%$ of the virial radius, the fitting formulae
used for the interpretation of the data may not be appropriate. For
this reason \citet{hayashi03} advocate comparing the observed rotation
curves directly against rotation curves measured from the
simulations. They conclude that the majority of the observed rotation
curves are adequately fit by CDM halos. Nevertheless a third of the
observed LSB rotation curves cannot be satisfactorily accounted
for. They postulate that the inconsistency might be caused by the
effects of halo triaxality on the motion of baryonic material and the
difference between circular velocity and gas rotation speed. This
highlights the main problem encountered by all dynamical analyses: The
approach has many complexities that prevent clear interpretation of
the observations.

These results motivate the search for an alternative method to measure
galaxy mass profiles, free of such ambiguities. Gravitational lensing
provides an attractive solution, primarily because the deflection
angle of a photon passing a massive object is independent of the
dynamical state of the deflecting mass. Therefore lensing is not
subject to any of the difficulties associated with dynamical
techniques, offering a straightforward approach based on well
established physics.

Strong lensing systems, where a background source is multiply imaged,
allow parameterised lens mass profiles to be constrained, by searching
for the best fit to the observed image positions. \citet{sand02}
enhanced this technique by incorporating extra constraints from the
velocity dispersion profile of the lens, and this has since seen
application to a number of systems \citep{treu02,koopmans03,sand04}.
Nevertheless, \citet{dal03} have criticised these results, arguing
that the tight constraints claimed were driven by prior assumptions
and that in general, more detailed modelling is required.

If the background source has extended structure, multiple arc
images or Einstein rings are formed. With high--resolution data, the
image will comprise a large number of resolution elements. Extended
sources therefore have the considerable advantage that they can
provide many more constraints on the lens mass profile compared to
images of point sources. A complete analysis of images of extended
sources requires modelling of the source surface brightness
distribution. The properties of both the source and the lens must
be adjusted to give the best fit to the observed ring. Additionally, a
proper solution to this inversion problem must also account for the
convolution of the image with the point spread function ({\em
psf}). \citet{warren03} (hereafter WD03) provide a summary of the
various approaches to this inversion problem that have been suggested.

The way in which the source is modelled can have far--reaching
effects. Because real sources have complicated structure, assuming an
over--simple analytic source surface--brightness profile can bias the
mass model solution, since in the minimisation the mass model will
attempt to compensate for the shortcomings of the source model. A
non--parametric form, for example where the source surface brightness
distribution is pixelised, overcomes this difficulty.

\citet{wallington96} describe a method that uses a pixelised source
surface brightness distribution. The solution is reached by searching
through the parameter space of the mass distribution and the source
surface brightness distribution to find the best fit, i.e. the
combination that produces the model image which, convolved with the
{\em psf}, minimises a merit function. The merit function is the summed
$\chi^2$ of the fit of the model image to the data, plus a term
proportional to the negative entropy in the source plane. The entropy
term is a regularisation term that prevents amplification of noise
due to the deconvolution and forces a smooth (and positive) solution
for the source. For the sake of efficiency, the method employs two
nested cycles. The outer cycle adjusts the lens mass model, while the
inner cycle adjusts the surface brightnesses of the source pixels, to
produce the best fit for the particular mass model. The method was
recently applied by \citet{wayth04} to the source modelled in this
paper, the Einstein ring 0047$-$2808.

In WD03, we presented a new method, termed `semi--linear', for solving
this inversion problem.  Algebraically the method is very similar to
the maximum--entropy method in that we simply replace the entropy term
with a linear regularisation term. However, the method is quite
distinct in application because it replaces the source minimisation
cycle with a single, linear matrix inversion.  This guarantees that
the best source fit is obtained for a given lens model and also speeds
up the inversion.

In this paper, we apply the semi--linear method to HST--WFPC2
observations of the Einstein ring 0047$-$2808. Our analysis builds
upon the work of \citet{wayth04}, who used the same data, and
investigated a range of single component mass models. One of these was
a model in which the mass follows the light, with a single variable,
the mass--to--light ratio ($M/L$). This model provided a poor fit, and
this motivates an analysis which includes a dark--matter halo, to
investigate what constraints the data provide on the amount of dark
matter, and the value of the inner slope of the density
profile. Accordingly, here we model the lens with two components, a
baryonic component, for which the mass follows the light, nested in a
dark halo. We show how the contribution from each component can be
separated to allow measurement of the baryonic $M/L$ and the inner slope
of the dark--matter mass profile. We compare this model against two
single--component models, the singular isothermal ellipsoid, and the
power--law ellipsoid.

In WD03 we included a discussion of the advantages and disadvantages
of regularising the inversion. Regularisation allows a small source
pixel size to be used, which is desirable, to extract fine detail of
the structure of the source. This is possible because regularisation
stabilises the inversion, suppressing the amplification of noise
associated with the deconvolution of the {\em psf}.  Nevertheless we
argued that regularisation is not reliable for quantitative work. By
the use of simulations we showed that in some circumstances the
regularised solution, while providing a satisfactory fit to the image,
produced a reconstructed source light profile that was inconsistent
with the input model. With real data it would be impossible to
identify such an inconsistency. A further problem with regularisation
is that, by smoothing the source light profile, it effectively reduces
the number of parameters fitting the source, by an amount that cannot
be quantified. Since the total number of degrees of freedom in the
problem is then unknown, it is impossible to correctly assess the
goodness of fit of the solution. This prevents a proper statistical
comparison between models (see additional comments on this point by
\citet{kochanek04}). For these reasons we have sought to
develop a stable inversion method that avoids regularisation, but
still makes maximum use of the information in the image. As we
demonstrate in this paper, the key is to recognise that the fixed
resolution of the image translates to variable resolution in the
source plane, so that a variable pixel size across the source plane is
required. A further advantage of unregularised solutions is that the
covariance matrix for all the parameters is easy to compute (WD03).

The layout of the paper is as follows: In the next section, we provide
an outline of the semi--linear method and describe its extension to
include adaptive source plane pixelisation. In Section
\ref{sec_analysis} we outline the HST data preparation, provide
details of the three lens models fitted, and describe our minimisation
procedure and the computation of the uncertainties. The results of the
fitting are presented in \ref{sec_results}. We find that the
dual--component model provides a significantly better fit than the
other two models, and we analyse the results for this model in more
detail. We provide a brief discussion and summary in Section
\ref{sec_discussion}

We adopt a cosmology with $\Omega=0.3$, $\Lambda=0.7$, and
$H_{\circ}=65$km\,s$^{-1}$\,Mpc$^{-1}$ throughout.

\section{The semi--linear reconstruction method}
\label{sec_method}

For a full description of the semi--linear inversion method, we refer
the reader to WD03. In this section we outline only the main features
of the method.

The inversion relies on the fact that both the source plane and the
image plane are pixelised. The manner in which the source plane is
pixelised is not restricted, allowing the concentration of smaller
pixels in regions where stronger constraints exist. Pixels in the
image plane are labelled by the index $j=1,J$. We use $d_j$ for the
surface brightness in pixel $j$ and $\sigma_j$ for its uncertainty.
For a fixed lens mass model, one can form the set of $I$ {\em
psf}-smeared images $f_{ij},j=1,J$ for each source pixel $i$ having
unit surface brightness. We may then pose the question: What set of
scalings $s_i$ are required for these images such that their
coaddition yields a {\em model image} which provides the best fit to
the {\em observed image}? These scalings $s_i$ are then the most
likely surface brightnesses of the source plane pixels, for the given
mass model.

For unregularised inversion, the merit function is:
\be
\label{eq_basic_chisq}
G=\chi^2=\sum_{j=1}^{J}
\left(\frac{\sum_{i=1}^{I}s_i f_{ij}-d_j}{\sigma_j}\right)^2.
\ee

The min$-\chi^2$ solution is a linear one and is given by:
\be
\label{eq_reg_chisq_soln}
\bsf{S}=\bsf{F}^{-1}\,\bsf{D}.
\ee
The vector $\bsf{S}$ holds the source pixel surface
brightnesses $s_i$ and the square matrix \bsf{F} and vector \bsf{D} are
defined by:
\be
\bsf{F}_{ik}=\sum_{j=1}^J f_{ij}f_{kj}/\sigma_j^2\, , \quad
\bsf{D}_i=\sum_{j=1}^J f_{ij}d_{j}/\sigma_j^2.
\ee

This linear inversion step, providing the min$-\chi^2$ fit for a given
mass model, is the heart of the semi--linear method. The full solution
proceeds by a non--linear search of the parameter space of the mass
model to find the best of the min$-\chi^2$ fits. Note that the lens
mass model typically requires only a small number of parameters
compared to the number of source pixels. Therefore the semi--linear
method gives a vast reduction in the overall size of the parameter
space that must be searched. In WD03 we explained how the full
covariance matrix for all the (mass+source) parameters is closely
related to $\bsf{F}$ for the best mass model, and is easily computed.

Regularisation may be implemented in the semi--linear scheme by
adding a linear term to the merit function, of general form
$G_L=\sum_{i,k}a_{ik}s_is_k$. The coefficients $a_{ik}$
depend on the type of regularisation used. The merit function becomes:
\be
\label{eq_G}
G=\chi^2+\lambda G_L
\ee
In this equation, $\lambda$ is a constant which weights the level of
regularisation. Increasing $\lambda$ produces a smoother source but
pushes $\chi^2$ away from the minimum achieved in the unregularised case.

Regarding the type of regularisation (see \citet{nr01} for a detailed
description), in WD03 we found that different schemes make rather
little difference to the reconstructed source. In this paper, we use
the so-called zeroth--order regularisation, where
$G_L=\sum_{i=1}^{I}s_i^2$. This makes no reference to the relative
locations of source pixels, greatly simplifying the practical
implementation of our sub--pixelisation scheme (Section
\ref{sec_adap_grid}).

With linear regularisation, the solution becomes
\be
\label{eq_unreg_chisq_soln}
\bsf{S}=\left(\bsf{F}+\lambda\bsf{H}\right)^{-1}\,\bsf{D}
\ee
where the elements of the matrix $\bsf{H}$ are given by:
\be 
  \label{eq:eqn}
  \bsf{H}_{ik}=\frac{1}{2}\frac{\partial{^2G_L}}{\partial{s_i}\partial{s_k}}.
\ee
For zeroth--order regularisation $\bsf{H}$ is simply the identity matrix.

If regularisation is implemented, the full covariance matrix for the
lens and source parameters can only be obtained by Monte Carlo
methods.

\subsection{Adaptive source plane grid}
\label{sec_adap_grid}

In this sub--section we describe our scheme for choosing a grid of
pixels of varying size across the source plane. The goal is to choose
a pixelisation which maximises the information in the reconstructed
source but maintains a stable inversion. In this way regularisation of
the inversion will not be required.

There are a number of ways in which a variable source pixel scale might
be implemented. We have tested a variety of methods, including one
which attempts to control inter--pixel statistical dependencies by
varying their size according to the covariance between adjacent
pixels. Our method of choice in this paper is, instead, to vary the source
pixel scale according to the magnification, since this determines how
strongly different areas of the source plane can be constrained. In
this way, the error on the surface brightness of each reconstructed
source pixel is more constant across the whole of the source plane.
This solves the problem we found in WD03, that it is hard to choose a
compromising single source pixel size. If the sampling of the source
plane is too high, regions of low magnification give a very noisy
reconstructed source image. Conversely, a grid of pixels
that is too coarse loses information and can give a bad fit to the
image, biasing the lens solution.

Before describing the method adopted, it is important to determine the
smallest suitable source pixel scale in a region of low magnification.
Because the inversion involves deconvolution, the source pixel size
should be no smaller than Nyquist sampling of the {\em psf} inverted
to the source plane.\footnote{In fact, in more detail, the minimum
pixel size depends on the S/N of the data and on the {\em psf} power
spectrum, so that for data of higher S/N it would be possible to use a
smaller pixel size than advocated here.} At the wavelength of
observation, 550nm, the resolution of a diffraction limited 2.4m
telescope is $0.06\arcsec$ FWHM. This is a misleading representation
of the HST-WFPC2 image quality for three reasons: 1) The HST {\em psf}
includes broad low--level wings that reduce information content at
small scales. 2) The core of the delivered {\em psf} is undersampled
by the $0.1\arcsec$ pixels of the WFPC2 Wide Field Camera. 3) The
final image is further degraded by the pixel scattering function
\citep{biretta}.

To compute the appropriate sampling, we used the {\em TinyTIM}
software \citep{krist} to create a highly sampled {\em psf}
image. This is the {\em psf} in front of the detector. This image was
then convolved with a square $0.1\arcsec$ pixel, and then convolved
further with the pixel scattering function. The realised image of a
point source may then be thought of as a $\delta-$function sampling on
the pixel grid of this convolved function, with noise added. To
account for the loss of information due to the broad wings of the {\em
psf}, rather than directly measure the FWHM of this pixel--convolved
{\em psf}, we measured the radius which encloses $70\%$ of the energy,
and then computed the FWHM of the Gaussian which contains $70\%$ of
the energy within the same radius. The result was a FWHM value of
$0.24\arcsec$. Therefore, in an image of low {\em S/N}, there is
little information at smaller scales than this value, suggesting that
the pixel scale of the reconstructed source should be no smaller than
$0.12''$ in regions of low magnification. It should be noted that the
sub--pixel dithering strategy used for our observations (\S3.1) only
improves the sampling and not the resolution of the data.

The magnification, $\mu$, gives the ratio of the area of the image of a
source plane pixel (summed over all copies) to the original source
pixel area. Roughly speaking, $\sqrt{\mu}$ represents the improvement in the
resolution in translating from the image plane to the source
plane. Therefore, one might expect that to maximise the information
in the reconstructed source, the source pixel area should scale
inversely with $\mu$. 

To implement such a scheme, a magnification map for a mass model close
to the final solution is needed. For this purpose we computed the best
fit lens model obtained with a regular source grid of pixel scale
$0.06\arcsec$. Using this model, the adaptive pixelisation starts with
an initial grid of source pixel scale $0.12\arcsec$. Depending on the
magnification, these pixels are split into 4 and some sub--pixels
further split into 4, resulting in a minimum source pixel scale of
$0.03\arcsec$. Naively, if the average magnification over a pixel
satisfies $\mu>4$, the pixel should be split, and if within a
sub--pixel $\mu>16$, the sub--pixel should be split. In reality the
resolution improvement is direction dependent, since source pixels are
not isotropically magnified, and so a more conservative criterion is
needed. This is also desirable because the magnification across a
pixel can vary rapidly. The splitting criterion refers to the average
magnification across a pixel, but the condition may not be true of
each of the sub--pixels into which the pixel is split. For these
reasons, instead of the factors 4 and 16 above, we introduce the `{\em
splitting factor}', $s$, such that a pixel is split if $\mu>s$ and a
sub--pixel is split if $\mu>4s$. The problem is thus reduced to
identifying the optimal value of $s$.  Clearly if the splitting factor
is large, only a few highly magnified pixels will be split. The source
pixels will then be too large to match all the detail in the image,
and information will be lost. If, on the other hand, the splitting
factor is small, in regions of low magnification they will oversample
the inverted {\em psf}, resulting in a very noisy reconstructed
source. 

\begin{figure*}[ht]
\epsfxsize=165mm
{\hfill
\epsfbox{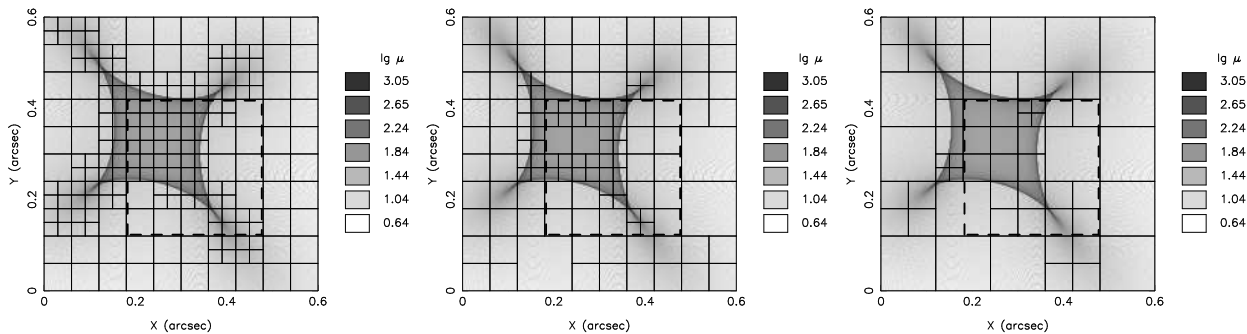}
\hfill}
\caption{Source plane pixelisations corresponding to splitting factors
of $s=$4, 9, and 14 from left to right. The grey scale shows
log(magnification) calculated from the best fit dual--component lens model
in Section \ref{sec_results1}. Heavy dashed box shows the $0.3''
\times 0.3''$ source plane size used in the reconstruction throughout
this paper.}
\label{pixels}
\end{figure*}

We determine an optimal value of $s$ empirically, by measuring
the improvement in the fit brought about by the splitting for
different values of $s$, successively reducing the value of $s$ to the
point at which no significant improvement is obtained.
In detail, starting with a large value of $s$, the source plane is
pixelised as described, and the best fit model is recomputed (see
Section \ref{sec_min_proc}). This gives a set of minimised lens
model parameters, the reconstructed source surface brightness
distribution and the value of $\chi^2$ for the fit.  The value of $s$
is then reduced, the procedure repeated, and the new value of $\chi^2$
computed. If the reduced value of $s$ increases the number of source
pixels by $\Delta I$, then this is the decrease in the number of
degrees of freedom. The change $\Delta\chi^2$ should therefore be
distributed as $\chi^2$ for $\Delta I$ degrees of freedom, since we
have simply increased the number of linear parameters in the fit. If
$\Delta\chi^2$ is significant by this test, the lower splitting factor
is accepted and the next lower value of $s$ is then tested. We set the
significance level at the conservative value of $1\%$, for the reasons
given above.

Figure \ref{pixels} illustrates the changing pixelisation as the
splitting factor is decreased successively through the values, 14, 9,
4, from right to left.  In each plot, the grey scale is the $\mu$ map.
We find that the middle value $s=9$ corresponds to the $1\%$
significance level chosen, for all three lens models described in
Section \ref{sec_model}. Comparison of the goodness of fit of each
model in Section \ref{sec_results1} is therefore carried out with an
adaptive source grid constructed using $s=9$.

In order to ensure a completely fair comparison between models, a
further effect must be taken into consideration. The chosen offset of
the source plane centre with respect to the centre of the lens caustic
structure can in principle bias the goodness of fit of one lens model relative
to another. Each of the three lens models tested has a slightly
different caustic shape.  A given source plane offset can result in a
more effective adaptive pixel grid for one lens model than another due
to fortuitous alignment of pixel edges relative to the lens
caustic. We deal with this effect, essentially, by including the source
plane offset in the minimisation; we perform a full lens $+$ source
minimisation at every point on a regular grid of offsets and take the
best overall fit. This is discussed further in Section
\ref{sec_min_proc}.

Our method of adaptively pixelising the source plane in this way
solves the problem noted by \citet{kochanek04} that plagues current
pixelised source based methods. Existing techniques use a regular
source grid and thereby rely on some form of regularisation to control
the behaviour of the reconstructed source in regions of low
magnification.  Regularisation smooths the source light profile,
reducing the effective total number of parameters and hence increasing
the number of degrees of freedom, by an amount that cannot be
quantified.  This is especially problematic when comparing different
lens models, as a fixed regularisation weight for one model generally
would not give the same increase in number of degrees of freedom for
another. In our scheme, the splitting factor has been chosen such that
the adaptively sized pixels extract maximum information from the lens
image without need of regularisation. Therefore, the number of degrees
of freedom of the fit is a well--defined number. This allows, firstly,
unambiguous assessment of whether a given model provides a
satisfactory fit to the data and secondly, unbiased comparison of
different model fits.

\section{Data and method of analysis}
\label{sec_analysis}

In this section we provide details of the observational data
(\ref{sec_data}) and the lens models considered (\ref{sec_model}).
We also discuss the practicalities of performing the minimisation
(\ref{sec_min_proc}) and explain the calculation of the uncertainties
(\ref{sec_model_errors}).

\subsection{HST observations}
\label{sec_data}

The data analysed here are the same data analysed by \citet{wayth04},
who give full details of the observations and data reduction. In
order to keep the current paper self--contained, the key elements are
outlined in this section.

The field of 0047$-$2808 was observed with HST's WFPC2 instrument in
the F555W filter over four orbits. We used the WFC, which has
$0.1\arcsec$ pixels. The chosen filter ensured that the strong
Ly$\alpha$ emission from the source star--forming galaxy at $z=3.595$
\citep{warren96} was detected, thereby enhancing the ring:lens flux
ratio. Observations were dithered using a $2\times2$ pattern with a
horizontal and vertical step size equal to $N+0.5$ pixels. At each of
the four dither positions, two exposures of 1200s were taken to aid
cosmic ray removal.

\begin{figure}[h]
\epsfxsize=60mm
{\hfill
\epsfbox{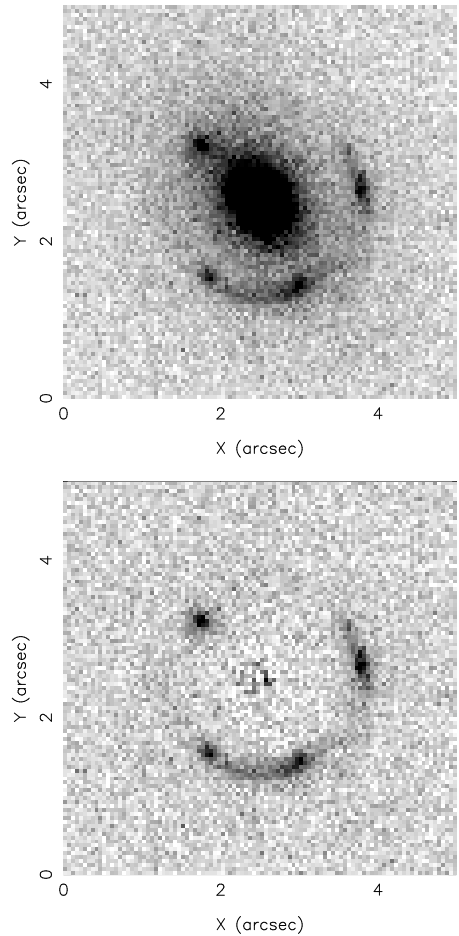}
\hfill}
\caption{{\em Top}: Interlaced HST image of $0047-2808$ with pixel
interval $0.05''$. {\em Bottom}: Interlaced image after subtracting
the image of the lensing galaxy. In this figure, the position angle of
the vertical axis is $27.44^{\circ}$ E of N.}
\label{hst_obs}
\end{figure}

After subtracting the background counts from each exposure and
eliminating cosmic rays, pairs of exposures at each dither position
were averaged. These four combined images were then used to form an
{\em interlaced} image with pixel interval $0.05''$. This image is
reproduced in Figure \ref{hst_obs}. In the figure the pixels are shown
with side $0.05''$, but this is only for presentational purposes. In
the analysis, we account for the true size of $0.1''$ by fitting
simultaneously to the four images that make up the interlaced image.

We constructed a noise frame for this image for the purposes of
measuring $\chi^2$, both for fitting the light profile of the lensing
galaxy and for the lensing analysis. Our Poisson estimate of the
pixel flux uncertainties allows for photon noise and readout noise,
and accounts for the removal of cosmic rays.

The image of the lensing galaxy was subtracted by fitting a S\'{e}rsic
profile \citep{sersic68} plus a central point source
\citep{wayth04}. The best fit was found by minimising $\chi^2$,
discounting an annular area containing the image of the lensed
source. In the fitting procedure profiles were convolved with a model
WFPC2 {\em psf} and the appropriate pixel scattering kernel. The lower
half of Figure \ref{hst_obs} shows the resulting Einstein ring after
the lensing galaxy has been subtracted. The residue at the centre of
the subtracted image is not significant, given the high counts at the
centre of the lens galaxy image. This is the image used in the lensing
analysis.

\subsection{Lens models}
\label{sec_model}

\citet{wayth04} used the same data analysed in the current paper to
test a range of commonly used lens mass models. Three models failed to
provide a satisfactory fit. These were: 1) a model in which the mass
profile follows the light profile, with a single free parameter, $M/L$,
2) the NFW model
$\rho(r)=\frac{\rho_s}{(r/r_s)(1+r/r_s)^{2}}$ \citep{nfw96}, with
$r_s=25$kpc, a value suggested by N--body simulations for galaxies of
this mass, 3) a model of a singular isothermal sphere with external
shear. In contrast, the singular isothermal ellipsoid (SIE) model, as
well as three other similar models, were found to provide 
satisfactory fits. Nevertheless, a general power--law model, of which
the SIE is a special case, gave a better fit than the SIE model, at a
marginally significant level.

The current paper follows on from \citet{wayth04}. Since the formalism
of the semi--linear method is different from the maximum--entropy
method applied by \citet{wayth04} (Section \ref{sec_intro}), we
repeat part of their analysis by fitting the SIE and power--law
models, confirming their conclusion that the power--law model is
preferred. We then go on to test a dual--component model, comprising
baryons and dark matter. Details of the various models are provided
below.

We use coordinates $x,y$ defined by the axes of the CCD array and
centred on the centre of the galaxy light distribution. The models
include mass components with elliptical surface mass densities
characterised by four parameters; the coordinates of the centre $x,
y$, the axis ratio $q=b/a$ (i.e. the ratio of the semi--minor axis to the
semi-major axis), and the orientation of the semi--major axis
$\theta$, defined as the angle measured counter--clockwise from the
vertical. The surface mass density profiles are parameterised in terms
of the ellipse coordinate $\xi$, defined by
$\xi^2=x^{\prime2}+y^{\prime2}/q^2$, where $x^\prime$, and $y^\prime$,
are coordinate axes aligned with the semi--major and semi--minor axes
of the ellipse.

For each mass model the components of the deflection angle vector
{\mbox{\boldmath $\alpha$}} are computed from the surface mass 
density profile $\Sigma(\xi)$ using the method of \citet{schramm90}, 
as parameterised by \citet{keeton01}; 
\be
\label{eq_schramm_defl}
(\alpha_1,\alpha_2)=(qx^{\prime}J_0(x^{\prime},y^{\prime}),qy^{\prime}
J_1(x^{\prime},y^{\prime}))
\ee
where
\be
\label{eq_defl_ang}
J_n(x^{\prime},y^{\prime})=\int^1_0\frac{\Sigma(\xi)/\Sigma_c}
{[1-(1-q^2)u]^{n+1/2}}{\rm d}u
\ee
and where $\Sigma_c$ is the critical surface mass density and
\be
\label{eq_eta_defn}
\xi(u)^2=u\left(x^{\prime2}+\frac{y^{\prime2}}{1-(1-q^2)u}\right).
\ee

The different models are tested against each other by comparing
the best fit values of $\chi^2$ in relation to the numbers of degrees
of freedom of the models. In all cases we eschew regularisation and
use the formalism previously described to select the optimal pixelisation
of the source plane so that the number of degrees of freedom is well
defined.

\subsubsection{SIE and singular power--law models}
\label{sec_single_pl}

The power--law model has a volume mass density profile of the form 
\be
\rho(r)=\rho_0 (r/r_0)^{-\alpha} 
\ee 
where $r_0$ is arbitrarily chosen. We allow the coordinates of the
centre of the mass to be offset from the centre of the light.  Adding
$\rho_0$ and $\alpha$ to the four parameters of the ellipse, the
power--law model has 6 parameters.  The SIE model is the special case
$\alpha=2$, and has 5 parameters.

\subsubsection{Baryons + dark matter halo model}
\label{sec_dual_model}

We assume that the projected surface mass density of the baryonic
component of the lens model follows the surface brightness
distribution of the lens galaxy. The baryons are therefore fixed in
shape as determined by the S\'{e}rsic $+$ point source profile fitted
to the image.  In our analysis, the baryonic contribution to the total
mass profile is determined by the rest--frame B--band baryonic $M/L$,
$\Psi$ (in units $h_{65}M_{\odot}/L_{B\odot}$), which is left as a
free parameter in the minimisation. (In converting from the {\em
F555W} filter to the rest--frame B--band, we have adopted the
$k-$correction and the band zero--point difference computed by
\citet{koopmans03} (see \citet{wayth04} for details).)  The lens
deflection angle due to the baryons is hence calculated from the
elliptical surface mass density of the fitted S\'{e}rsic profile
which, in units of $M_{\odot}/\Box''$, is
\be
\label{eq_sersic_smd}
\Sigma_b(\xi_b) = \Psi L_{1/2} \exp\{-5.90[(\xi_b/1.09'')^{0.32}-1]\}
\ee
and the mass of the central point source,
\be
M_{p}=3.09\times 10^9 \,h_{65}^{-2}\,L_{B\odot} \Psi,
\ee
expressed here in units of $M_{\odot}$. 
The fitted half light luminosity of the S\'{e}rsic profile
is $L_{1/2}=(1.99\pm0.09)\times 10^9 h_{65}^{-2} L_{B\odot}/\Box''$.
The axis ratio measured from the
light distribution is $q_b=0.69$ \citep{wayth04}. Since the 4
parameters of the ellipse are the measured values for the S\'{e}rsic
fit, the baryonic component of the mass model has a single free
parameter $\Psi$. Note that the deflection angles for this model need
only be computed once, and then scaled by $\Psi$ as $\Psi$ is varied
in searching for the best--fit mass model.

For the dark matter halo, we choose a generalised NFW model \citep{nfw96}
which allows for a variable central density profile slope $\gamma$.
This has a volume mass density profile given by
\be
\rho(r)=\frac{\rho_s}{(r/r_s)^{\gamma}(1+r/r_s)^{3-\gamma}}.
\ee
Here, $\rho_s$ is the halo normalisation and $r_s$ is a scale radius.
To convert this to a projected surface mass density profile, we
integrate along the line of sight following the prescription of
\citet{keeton01}. This integration, which must be evaluated
numerically, gives a radially symmetric surface mass density
profile. As with the S\'{e}rsic profile, ellipticity is introduced
by replacing the radial coordinate with an ellipse coordinate
$\xi_h$ that has an associated axis ratio $q_h$.

In total there are 6 free parameters for the dark--matter mass
profile, $x_h, y_h, q_h, \theta_h, \rho_s, \gamma$. We hold the scale
radius fixed at $r_s=50 h_{65}^{-1}$kpc $(\cong 8'' \,\, @ \,
z=0.485)$ to match that expected from simulations by
\citet{bullock01} for a galaxy of similar mass and redshift as the
lens galaxy in $0047-2808$. In Section \ref{sec_model_errors} we
discuss the effect of changing $r_s$, although since it is much larger
than the radius of the ring traced by the observed images $r=1.16''$,
this effect is small.

The overall deflection angle at any point in the lens plane resulting
from the combined effect of the baryonic and halo mass is simply given
by the addition of the separate deflection angles due to the point
mass, the S\'{e}rsic profile and the generalised NFW halo. 

The combined model, baryons and dark matter, has 7 parameters.

\subsection{Minimisation procedure}
\label{sec_min_proc}

A full minimisation for a given lens model involves three nested
processes.  The innermost process is the linear inversion step
explained in Section \ref{sec_method}, giving a reconstructed source
and $\chi^2$ from equation (\ref{eq_basic_chisq}) for a trial set of
lens model parameters.  The middle process minimises the lens model
parameters and for this we use Powell's method \citep{nr01}. Finally,
at the outermost level, we step through a grid of source plane offsets
to address the effect discussed in Section \ref{sec_adap_grid}, that
fortuitous alignments of the source pixel grid with the lens caustic
can bias the fit. We step through a $10 \times 10$ grid of offsets of
size $0.006''$, corresponding to a tenth of a medium-sized pixel in
Figure \ref{pixels}. The best overall fit is taken as the lowest value
of $\chi^2 - N_{dof}$.

In computing the images of each source pixel, we sub-grid each image
plane pixel into a $4 \times 4$ array of sub--pixels. Rays are traced
back from the image plane to the source plane via each image plane
sub--pixel. The sub--gridding was chosen to ensure a smooth $\chi^2$
surface, necessary for reliable minimisation.

For the dual--component model we calculate joint confidence regions in
the $\gamma-\Psi$ plane by marginalising over the remaining 5
parameters, at regular grid points spanning this plane. At every point
in the $\gamma-\Psi$ plane, the minimisation is initialised with the
halo centred on the S\'{e}rsic centre and possessing the same
orientation and ellipticity as the visible light. The normalisation of
the halo must be initialised by a fitted function as explained below.
Satisfactory convergence is reliably obtained by setting an
arbitrarily small tolerance and terminating minimisation once 200
iterations have been performed.

We find at all points in the parameter space that the variation of
$\chi^2$ with halo normalisation has three minima near the correct
lens solution. One is the correct solution, defined such that the
reconstructed source surface--brightness distribution is most
compact. In addition there are two local minima corresponding to under
and over magnifications. In the under magnified case, the
reconstructed source surface brightness distribution is a smaller,
slightly distorted version of the observed ring image. In the over
magnified case, the source resembles a small ring image but inverted
such that every pixel in the ring has been reflected in a plane
running through the ring centre and perpendicular to the pixel's
radial vector.  Both under and over magnified source reconstructions
can be rejected on the grounds that they produce extraneous images
when lensed to the image plane. To prevent convergence to an incorrect
local minimum, we initialise the halo normalisation $\rho_s$ to a
value which is estimated to lie close to the correct solution.  This
initial value is set by a fitting function derived from a simplified
analysis in which only $\rho_s$ is allowed to vary across the
$\gamma-\Psi$ plane.

\subsection{Modelling uncertainty}
\label{sec_model_errors}

The uncertainty on the set of reconstructed lens model parameters is
determined from the prescription in WD03. This is calculated by
inverting the curvature matrix for all parameters, including the
source pixels, to obtain the full covariance matrix. The uncertainty
from the fit for a given parameter is then just the relevant diagonal
term in this matrix. For the dual--component model, we determine the
error on $\gamma$ and $\Psi$ directly from their marginalised $\chi^2$
contours. The total error that we quote for $\Psi$ includes an extra
contribution from the only source of significant uncertainty in the
Sersic profile; the parameter $L_{1/2}$ (see \ref{sec_dual_model}).

A final source of error in our dual--component model stems directly
from the uncertainty on the scale radius $r_s$ in the halo component.
\citet{dal03} argue that $r_s$ should be left as a free parameter in
the minimisation. We have opted to hold $r_s$ at the value of $50
h_{65}^{-1}$kpc as expected from simulations by \citet{bullock01} and
search for a solution in the context of this model. Our data can only
weakly constrain $r_s$ which has the advantage that our results do not
depend sensitively on its value. We find that a 10\% change in $r_s$
produces a $\sim 1\%$ change in the minimised $M/L$ and a negligible
change in $\gamma$. This error is not included in the final error
budget.

\section{Results}
\label{sec_results}

This section is divided into two halves. In the first, Section
\ref{sec_results1}, the three lens models are compared. In the second,
Section \ref{sec_results2}, the dual--component model is considered in
more detail and using this model, we reconstruct the source surface
brightness distribution.

\subsection{Comparison of Models}
\label{sec_results1}

In this section, we compare the performance of the three models,
1. SIE, 2. power-law, and 3. dual--component models, in fitting the
observed ring image.  We also compare our findings with the analysis
by \citet{koopmans03} who analysed 0047$-$2808 using a method
combining dynamical measurements and lensing.

All reconstructions in this section are unregularised. As explained in
section \ref{sec_adap_grid} this is to allow unbiased comparison of
models. The $\chi^2$ is evaluated in an annular masked region shown in
the bottom left panel of Figure \ref{recon_source}.  The mask was
designed to ensure that it includes the image of the entire
source plane, with minimal extraneous sky. This means that only
significant image pixels are used in the fit, making $\chi^2$ more
sensitive to the model parameters.

Table \ref{tab_cf_chisq} summarises the results, listing the best--fit
value of $\chi^2$ and the number of degrees of freedom (NDF) of that
fit. Recall that the NDF depends not only on the number of parameters
of the mass model, but also on the exact source pixelisation used,
which in turns depends on the structure of the caustics.

\begin{table}
\centering
\small
\begin{tabular}{|c|c|c|}
\hline
Model & $\chi^2_{min}$ & NDF \\
\hline
SIE & 1156.2 & 1247 \\
power--law & 1157.7 & 1255 \\
baryons+halo & 1161.4 & 1269 \\
\hline
\end{tabular}
\normalsize
\caption{Performance of singular isothermal ellipsoid (SIE),
power--law, and dual--component models, in terms of minimum $\chi^2$
and number of degrees of freedom.}
\label{tab_cf_chisq}
\end{table}

\begin{table}
\centering
\small
\begin{tabular}{|c|c|c|c|c|}
\hline
Parameter & Minimised Value & $1\sigma$ error in fit \\
\hline
$\rho_s$ & $3.27\times10^6$ & $0.22\times10^6$\\
$(x_h,y_h)$ & $(0.070'',0.026'')$ & $(0.004'', 0.005'')$ \\
$q_h$ &  0.820 & 0.020\\
$\theta_h$ &  $41.7^{\circ}$ & $1.5^{\circ}$\\
\hline
\end{tabular}
\normalsize
\caption{Remaining minimised parameters for halo in the dual--component model. 
Reading from top to bottom, parameters are; normalisation in units of
$h^2_{65}M_{\odot}$kpc$^{-3}$, offset from light centre, axis ratio, and
orientation of semi--major axis from vertical in counter--clockwise
direction. The $1\sigma$ error is derived from the full covariance
matrix.}
\label{tab_min_model}
\end{table}

\subsubsection{SIE and power--law models}

For the SIE model, $\alpha=2$, the best fit gives
$\chi^2_{min}=1156.2$ for 1247 degrees of freedom. The power--law
model gives $\chi^2_{min}=1157.7$ with 1255 degrees of freedom, and a
measured slope $\alpha=2.11 \pm 0.04$. The increase in the NDF is
because fewer pixels are used to tessellate the source plane.
Comparing the power--law to the SIE model, the increase in $\chi^2$ of
$\Delta \chi^2 = 1.5$, only, for an increase of 8 degrees of freedom
differs from the expectation of $\Delta \chi^2\sim8$ at a significance
level of $99.3\%$. The power--law model therefore is a significant
improvement over the SIE model. This is reflected in the measured
value of $\alpha=2.11$ which is inconsistent with $\alpha=2$ at the
$2.7\sigma$ significance level.

\citet{wayth04} compared the same two models, using a maximum--entropy
inversion method. This method entails regularisation of the
inversion. They applied minimal regularisation (such that the
inversion amounts to the non--negative min--$\chi^2$ solution), in
order to minimise the uncertainty in the change in the NDF. They found
that the power-law model gives a better fit than the SIE, at an
associated significance level of $96\%$. They found
$\alpha=2.08\pm0.03$ for the power-law model as well as orientations
and ellipticities for both models consistent with our findings.  These
results are in good agreement with ours, vindicating their approach
for dealing with the problem of the uncertainty of the NDF.

\subsubsection{Baryons + dark matter halo model}

With the dual--component model, we obtain $\chi^2_{min}=1161.4$ for
1269 degrees of freedom. Comparing this against the best fitting
singular power--law model gives an increase in $\chi^2$ of $\Delta
\chi^2 = 3.7$, only, for an increase of 14 degrees of freedom.  This
small increase in $\chi^2$, differs from the expectation of $\Delta
\chi^2\sim14$ at a significance level of $99.7\%$, demonstrating that
the dual--component model provides a significantly better fit.

\begin{figure}[htb]
\epsfxsize=65mm
{\hfill
\epsfbox{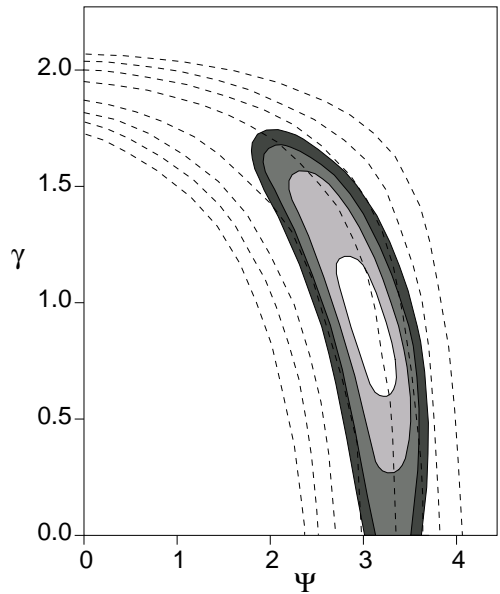}
\hfill}
\caption{One--parameter confidence regions on inner slope and baryonic
$M/L$: 68\%, 95\%, 99\% \& 99.9\%. Dashed lines are same confidence
regions from analysis by \citet{koopmans03}. }
\label{semi-linear_conts}
\end{figure}

Figure \ref{semi-linear_conts} shows the $\chi^2$ contours in the
$\gamma-\Psi$ plane, marginalised over the remaining 5 parameters.
The grey shaded regions give the 68\%, 95\%, 99\% \& 99.9\%
one--parameter confidence limits.  We obtain an inner slope of
$\gamma=0.87^{+0.69}_{-0.61}$ to 95\% confidence (or a limit of
$\gamma<1.74$ to $99.9\%$ confidence) and a $M/L$ of
$\Psi=3.05^{+0.53}_{-0.90} \,h_{65}\,M_{\odot}/L_{B\odot}$ to 95\%
confidence, including the photometric error (or
$\Psi=3.05^{+0.78}_{-1.30} \,h_{65}\,M_{\odot}/L_{B\odot}$ to $99.9\%$
confidence). The remaining five minimised lens model parameters are
provided in Table \ref{tab_min_model}.

The dashed lines in Figure \ref{semi-linear_conts} show the same
confidence levels obtained by \citet{koopmans03} in analysis of the
same HST data as in this paper. Their result used the total mass
enclosed by the Einstein ring and in addition, the measured velocity
dispersion profile of the lens galaxy, as constraints. Clearly, our
semi--linear method, in using all the information contained within the
ring image, significantly reduces the uncertainty in the lens
model. Overall the two results are in very good agreement, bearing in
mind that our result comes from a purely gravitational lens analysis,
while theirs is primarily a dynamical analysis, supplemented by the
size of the Einstein ring.

With reference to the other minimised parameters given in Table
\ref{tab_min_model}, we conclude that; 1) the centre of the baryons is
closely aligned with the halo centre, 2) the halo, with
$q_h=0.82\pm0.02$, is significantly rounder than the stellar component
of the galaxy, with $q_b=0.69\pm0.01$, 3) there is a significant
difference between the baryonic orientation of $\theta_b=35.0\pm0.7$
and that of the halo $\theta_b=41.7\pm1.5$ .

\subsection{Further Analysis}
\label{sec_results2}

In the previous section we established that the baryon+dark matter halo
model provides a significantly better fit to the observed ring
compared to single component models. We now analyse this model in more
detail.

\subsubsection{Baryonic mass \& light}

As Figure \ref{semi-linear_conts} shows, we have been able to
constrain the baryonic $M/L$ without the need for dynamical
measurements. This is the first time a pure lensing analysis of a
single system has measured the baryonic $M/L$ directly. Being a lens
estimated quantity, this is free of the uncertainties associated with
dynamical methods (see Section \ref{sec_intro}).

\citet{koopmans03} show that the lens galaxy in 0047$-$2808 is offset
from the local fundamental plane by a factor 0.37dex. This value is in
close agreement with the expected passive evolution of this galaxy,
estimated from population synthesis models matched to the measured
$V-I$ colour \citep{wayth04}. Correcting by this factor, our derived
$M/L$ $\Psi=7.1$ is remarkably similar to the local average value for
the centres of ellipticals of $7.3\pm
2.1\,h_{65}\,M_{\odot}/L_{B\odot}$ \citep{gerhard01,treu02},
determined dynamically. Either this is a coincidence, or it is an
indication that the various elements going into this comparison,
i.e. our lensing analysis, the local dynamical analysis, the
population synthesis models, and the assumption of passive evolution,
are all quite accurate.

\subsubsection{Halo and baryonic fractions}

We calculate the fractional contribution of the baryons to the total
projected mass by integrating the projected mass distribution of each
component inside a circular aperture of radius $1.16''$ placed at the
lens centre (the offset between components is small enough to
disregard).  Within this aperture, the baryons account for
$65^{+10}_{-18}$\% (95\% CL) of the total projected mass.

The fraction of dark matter inside a sphere of the same radius is
obtained by deprojecting both surface mass density profiles. Because
lensing measures only projected mass, we have no information regarding
its distribution along the line of sight and so this must be
assumed. Our approach is to first calculate a circularly averaged
surface density profile for each component and then deproject assuming
spherical symmetry.  Deprojection is carried out using the approach
given by \citet{binney87}.

\begin{figure}[h]
\epsfxsize=70mm
{\hfill
\epsfbox{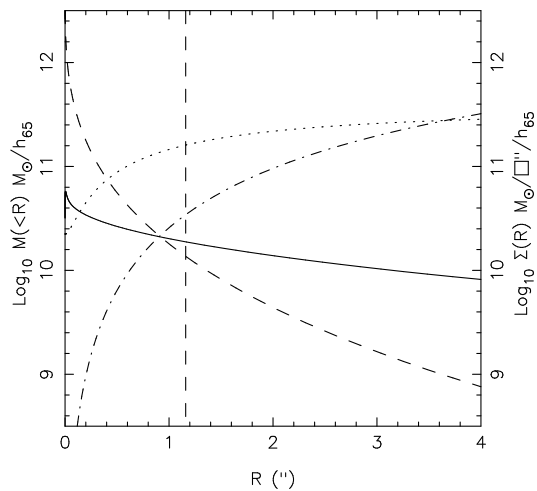}
\hfill}
\caption{Circularly averaged projected surface mass density profile of halo
(solid) \& baryons (dashed), and cumulative 3--dimensional mass profile
of halo (dot--dash) and baryons (dotted).  Vertical dashed line is mean
radius $1.16''$ of the ring traced by the lensed images. }
\label{cuml_mass}
\end{figure}

We find that the baryons account for $84^{+12}_{-24}$\% (95\% CL) of
the total mass within a sphere of radius $1.16''
(7.5h_{65}^{-1}$kpc).  Figure \ref{cuml_mass} plots both the circularly
averaged surface mass density profile and the cumulative deprojected
mass profile for the halo and baryons.  In the case of the baryons,
the point mass is included.

\subsubsection{Source reconstruction}
\label{sec_src_recon}

\begin{figure*}[!htb]
\epsfxsize=165mm
{\hfill
\epsfbox{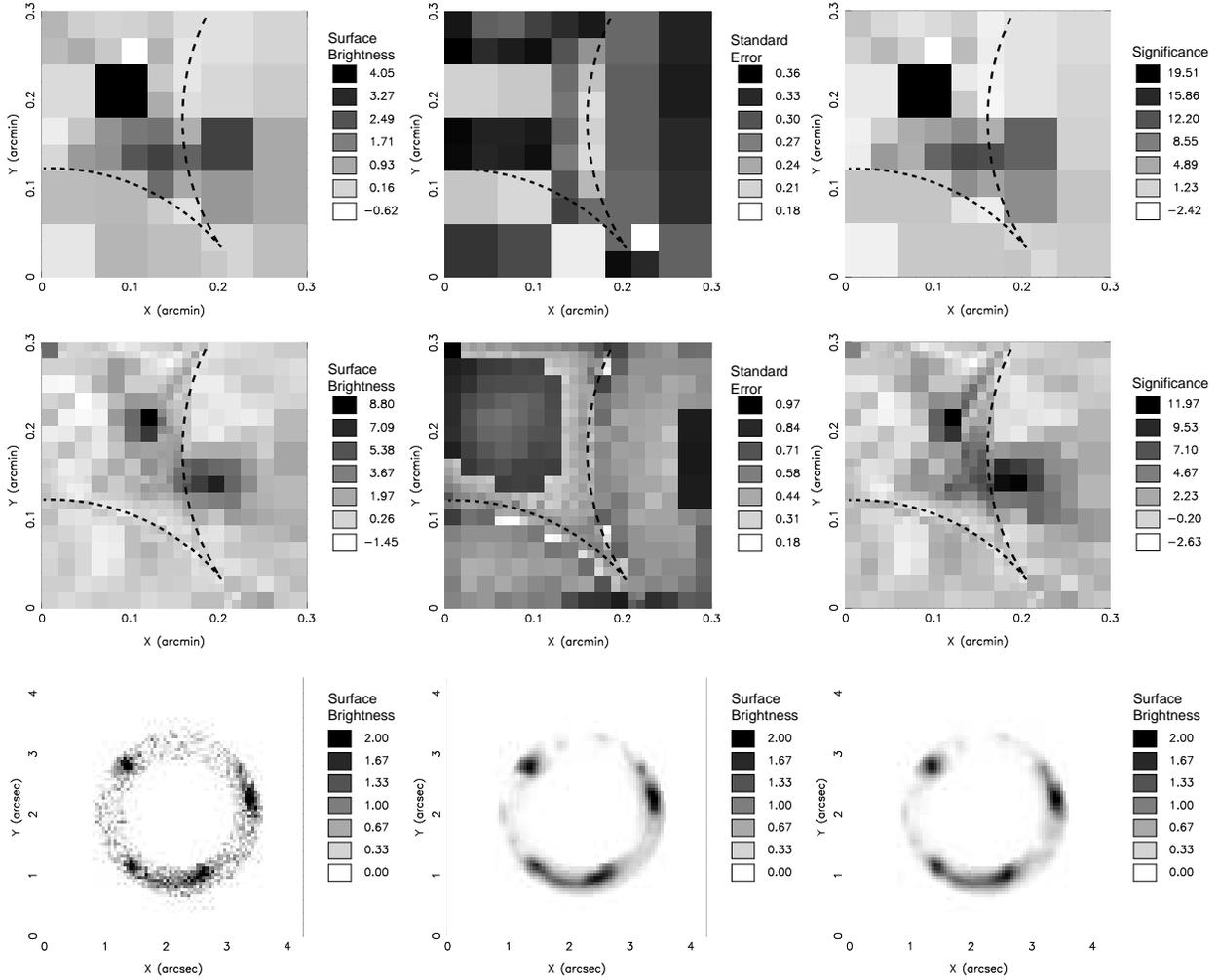}
\hfill}
\caption{Reconstructed source from best fit lens model (caustic shown
by dashed heavy line). {\em Top left:} Unregularised reconstructed
source surface brightness distribution obtained in Section
\ref{sec_src_recon}.  {\em Top middle:} Standard errors on
unregularised source pixels. {\em Top right:} Significance of
unregularised source. {\em Middle left:} Regularised source from best
fit lens model. {\em Middle:} Standard errors on regularised source.
{\em Middle right:} Significance of regularised source. {\em Bottom
left:} Masked observed ring image. {\em Bottom middle:} Lensed image
of unregularised source.  {\em Bottom right:} Lensed image of
regularised source}
\label{recon_source}
\end{figure*}

The reconstructed source and image for the best-fit dual component
model are shown in Figure \ref{recon_source}. The top row of this
figure shows the unregularised result.  Note the correspondence
between the pixelisation in these three panels, and that in the middle
panel of Figure \ref{pixels}. The top--left panel (side $0.3\arcsec$)
shows the reconstructed source, and the bottom middle panel (side
$4.2\arcsec$) is the image of this source, convolved with the WFC {\em
psf}. This is the best--fit to the actual image shown in the bottom
left--hand panel. The uncertainties on the source pixel surface
brightnesses are shown in the top middle panel, and the corresponding
significance map (ie. surface brightness divided by standard errors)
is shown in the top right--hand panel. Note that, in contrast with
similar maps shown in WD03, where the uncertainties were noticeably
smaller inside the caustic, here they are more uniform across the
source plane, as a consequence of the pixel size varying with the
magnification.

The limited source plane resolution allowed for these data by our
adaptive pixelisation scheme is a consequence of the relatively large
WFC pixels, and the relatively low S/N of the image. Nevertheless
there are clearly two areas in the source plane, of high significance,
where the source flux is concentrated, one on either side of the
caustic. This is the same conclusion reached by \citet{wayth04}. A
better sampled image of high S/N would allow formation of a clearer
picture of the nature of the source galaxy. Obviously, the source
light profile does not follow a simple parametric form. This means
that attempting to model 0047$-$2808 by forming model images using a
simple single source would bias the fitted model parameters.

Purely for the purposes of visualisation, we also performed a
regularised inversion. The results are provided in the middle row
of Figure \ref{recon_source}. For this, the regularisation weight,
$\lambda$, was chosen to make $\chi^2$ and the regularisation term in
equation (\ref{eq_G}) contribute equally to the figure of merit
$G$. Setting the minimum allowed source pixel size to $0.01'' \times
0.01''$, we followed the same procedure, set out in Section
\ref{sec_adap_grid}, to select the splitting factor.  The clarity of
the source is somewhat improved.  The outer source appears to be
extended, and to straddle the caustic. The significance plot,
right--hand panel, middle row, identifies both source components as
being highly significant. The reconstructed image, bottom right--hand
panel, is somewhat smoother, as expected.

\section{Discussion and summary}
\label{sec_discussion}

One of the main goals of this paper was to investigate the extent to
which a pure gravitational lens analysis of the image of an extended
source, using all the information in the image, could constrain the
inner slope of the dark matter density distribution in the lens
galaxy. Applying the method to the lens $0047-2808$,
we have succeeded in shrinking the uncertainties considerably,
compared to the lens+dynamical analysis of \citet{koopmans03}, which
used only the positions of the four brightest peaks in the image as
lens constraints. Our measurement of the inner slope of the dark
matter halo of $\gamma=0.87^{+0.69}_{-0.61}$ (95\% CL) is consistent
with the cuspy prediction of the CDM model. Nevertheless, we find that
the dark matter makes only a minor contribution to the total mass
within a spherical radius equal to the Einstein radius of the lens.
There is mounting evidence from both dynamical and lensing methods
that this is fairly typical of bright and intermediate--luminosity
early--type galaxies.  For example, \citet{romanowsky03} compared the
measured motions of planetary nebulae around three nearby ellipticals,
out to several effective radii, with dynamical models without dark
matter and found satisfactory agreement.  A similar, more
quantitative, conclusion was reached by a statistical analysis of 22
multiply--imaged quasars, by \citet{rusin03}. Although a single
multiply--imaged quasar is not useful for constraining dual--component
models, by assuming a fixed ratio of dark matter to baryons (within
two effective radii), the same power--law slope $\gamma$ for the dark
matter in all systems, and by invoking a relation between $\Psi$ and
galaxy luminosity, \citet{rusin03} were able to derive useful
constraints on dual--component models. By fixing the inner slope to
the NFW value $\gamma=1$, they find that the baryons account for
$78\%\pm10$ of the projected total mass within two effective radii,
average $7h_{65}^{-1}$kpc. This is consistent with our measurement of
$65\%^{+10}_{-18}(95\% CL)$, within $7.5h_{65}^{-1}$kpc, again in
2D. Note that we are able to achieve similar constraints from a single
system, with fewer assumptions. This highlights the usefulness of
images of extended sources.

As Figure \ref{cuml_mass} shows, the baryons are more concentrated
than the dark matter in this lens galaxy, and dominate the mass
distribution within the region of the image.  The baryons will alter
the shape of the dark matter halo, predicted by the pure dark matter
simulations, in a non--trivial way, dependent on the history of
assembly of the various components, the sequence of star formation,
and the extent to which gas is blown out of the galaxy by winds. A
simple treatment for estimating the influence of baryons is the so
called `adiabatic approximation' of \citet{blumenthal86}. In this
approximation, the expected profile of a collapsed halo can be
estimated from its initial profile and the initial and collapsed
baryonic profiles, assuming the halo adiabatically contracts. By
applying this in reverse, the initial halo profile can be estimated
from the collapsed profile as determined from a two component model
such as ours. This initial profile can then be compared directly with
the pure dark matter CDM simulations. \citet{treu02} used this reverse
method on their two component model of the lens system
MG2016$+$112. They found that the inner slope of the initial halo can
be substantially shallower than the measured collapsed slope. If this
is a valid approximation, then a similar effect would be expected for
0047$-$2808. This result is of interest for the case where the
measured slope $\gamma$ is already significantly smaller than the
predicted value, $\gamma\sim1$, since it acts to widen the
discrepancy. But given the simplifications involved, and therefore the
uncertainty in the magnitude of the effect, it is of limited interest
in the present case where our best--fit value is consistent with the
CDM value.

For our dual component model we found that the dark halo is offset in
orientation from the baryons by $6.7^{\circ}\pm1.7$, and is rounder,
with a significant difference in axis ratio of $\Delta
q=0.13\pm0.02$. Our model is a simple one, given the relatively low
{\em S/N} of the image, and does not include external
shear. Nevertheless, the galaxy does not lie in a dense environment,
as evidenced by the image provided in \citet{warren99}. There would be
some degeneracy in a solution including shear, between the amount of
shear and the orientation offset (see \citet{kks97} for a
discussion). A deeper image, with smaller pixels, would justify a more
sophisticated analysis. To gauge the potential improvement, we have
undertaken extensive simulations of observations of 0047$-$2808 with
the HST Advanced Camera for Surveys. Using the High Resolution Channel
over 10 orbits, with application of the semi--linear method we
anticipate a reduction in the error on $\gamma$ by a factor of $\sim
5$, sufficient to strongly test the CDM expectation. The improvement
is a consequence of the high throughput of ACS, and especially the
smaller pixel size.

We conclude with a summary of the main points of the paper:

\begin{enumerate}
\item We have extended the semi--linear method of WD03 for inverting
gravitational lens images of extended sources, to include adaptive
pixelisation of the source plane. We have identified a method for
tessellating the source plane that applies an objective criterion for
sub--dividing pixels, which maximises the information about the source
extracted from the image, and is also stable. Because of this the
inversion does not require regularisation. This eliminates the problem
that with regularised inversion the number of degrees of freedom is
ill defined. Proper statistical comparison of different mass models is
thereby enabled.

\item We have applied our semi--linear method to HST-WFPC2
observations of the Einstein ring system 0047$-$2808 to determine the
lens galaxy mass profile. We confirm the result of \citet{wayth04}
that a power--law model, $\alpha=2.11\pm0.04$ produces a significantly
better fit than the single isothermal ellipsoid model. Furthermore our
analysis shows that a dual--component model, comprising a baryonic
S\'{e}rsic profile + point mass nested in a dark matter generalised
NFW halo, gives a significantly better fit ($3.0 \sigma$) to the data
than the best power--law model. We demonstrated that using 100\% of
the information contained in the Einstein ring image, with an adaptive
source plane pixel scale, provides significantly better constraints
than the modelling of \citet{koopmans03}, who used the measured radial
variation of the stellar velocity dispersion, plus the lens
constraints provided by the positions of the four brightest peaks in
the ring.

\item For the dual--component model, we find that the
baryonic component has an unevolved rest--frame B--band $M/L$ of
$3.05^{+0.53}_{-0.90} \,h_{65}\,M_{\odot}/L_{B\odot}$. This
$M/L$ was obtained without any dynamical measurements and is therefore
not subject to the usual uncertainties associated with this
approach. The errors quoted here include the photometric uncertainty.
Evolving this value to zero redshift gives a result
consistent with the dynamically--measured $M/L$ in the centres of
nearby ellipticals. 

\item The measured inner slope of the dark--matter halo is
$\gamma=0.87^{+0.69}_{-0.61}$ (95\% CL), consistent with the
predictions of CDM simulations.

\item We find that the baryons account for $65^{+10}_{-18}$\% (95\% CL) of
the total projected mass or, assuming spherical symmetry,
$84^{+12}_{-24}$\% (95\% CL) of the total deprojected mass
within a radius of 1.16'' ($7.5h_{65}^{-1}$kpc) traced by the ring. 

\item We find that the dark--matter halo is significantly misaligned
  with the stellar light, and also is significantly rounder.

\item The reconstructed source surface brightness distribution shows
two distinct source objects in agreement with the findings of
\citet{wayth04}. This highlights the need for non--parametric sources
to obtain unbiased lens mass profiles.

\end{enumerate}

{\bf Acknowledgements}

We would like to thank Paul Hewett, Geraint Lewis, Leon Lucy, and
Randall Wayth for helpful discussions. SD is supported by PPARC.

\end{document}